\documentclass[11pt]{article}
\usepackage[letterpaper, hmargin=1in, vmargin=1.25in]{geometry}

    \renewcommand{\baselinestretch}{1.2}
  \renewcommand{\arraystretch}{1.1}
\begin{document}

 \title{Note on fast division algorithm for polynomials\\ using Newton iteration }

 \author{Zhengjun Cao\,$^*$, \quad  Hanyue Cao \\
 {\small Department of Mathematics, Shanghai University, Shanghai,
  China.} \\
  \textsf{$^*$\ \textsf{caozhj@shu.edu.cn}}  }

 \date{}\maketitle

\begin{abstract}

 The classical division algorithm for polynomials  requires $O(n^2)$ operations for inputs of size $n$. Using reversal technique and Newton iteration, it can be improved to $O(\mbox{M}(n))$, where $\mbox{M}$ is a multiplication time.
 But the method requires that the degree of the modulo,  $x^l$,  should be the power of $2$. If $l$ is not a power of 2 and $f(0)=1$,  Gathen and Gerhard  suggest  to compute the inverse, $f^{-1}$, modulo
  $  x^{\lceil l/2^r\rceil}, x^{\lceil l/2^{r-1}\rceil}, \cdots, x^{\lceil l/2\rceil}, x^ l$, separately.
 But they did not specify the iterative step.   In this note, we show that the  original Newton iteration formula can be directly used to
 compute $f^{-1}\, \mbox{mod}\,  x^{l}$ without any additional cost, when $l$ is not a power of 2.

\end{abstract}

\section{Introduction}

Polynomials over a field form a Euclidean domain. This means that for all $a, b$ with $b\neq 0$ there exist unique $q, r$ such that $a=qb+r$ where deg$\,r<$deg\,$b$. The division problem is then to find $q, r$, given $a, b$.   The classical division algorithm for polynomials  requires $O(n^2)$ operations for inputs of size $n$. Using reversal technique and Newton iteration, it can be improved to $O(\mbox{M}(n))$, where $\mbox{M}$ is a multiplication time. But the method requires that the degree of $x^l$ should be the power of $2$. If $l$ is not a power of 2 and $f(0)=1$,  Gathen and Gerhard \cite{GG03}  suggest  to compute the inverse, $f^{-1}$,  modulo
  $  x^{\lceil l/2^r\rceil}, x^{\lceil l/2^{r-1}\rceil}, \cdots, x^{\lceil l/2\rceil}, x^ l$, separately.
But they did not specify the iterative step.   In this note, we show that the  original Newton iteration formula can be directly used to compute $f^{-1}\, \mbox{mod}\,  x^{l}$  without any additional cost, when $l$ is not a power of 2.  We also correct an error in the cost analysis \cite{GG03}.

\section{ Division algorithm for polynomials using Newton iteration }

The description comes from Ref.[1].

Let $D$ be a ring (commutative, with 1) and $a, b\in D[x]$ two polynomials of degree $n$ and $m$, respectively. We assume that $m\leq n$ and that $b$ is monic. We wish to find polynomials $q$ and $r$ in $D[x]$ satisfying $a=qb+r$ with deg$r<$deg$b$ (where, as usual, we assume that the zero polynomial has degree $-\infty$). Since $b$ is monic, such $q, r$ exist uniquely.

Substituting $1/x$ for the variable $x$ and multiplying by $x^n$, we obtain
$$ x^na\left(\frac 1 x\right)=\left(x^{n-m}q\left(\frac 1 x\right)\right)\cdot \left(x^mb\left(\frac 1 x\right)\right)+x^{n-m+1}\left(x^{m-1}r\left(\frac 1 x\right)\right) \eqno(1) $$
We define the reversal of $a$ as $\mbox{rev}_k(a)=x^ka(1/x)$. When $k=n$, this is the polynomial with the coefficients of $a$ reversed, that is, if $a=a_nx^n+a_{n-1}x^{n-1}+\cdots+a_1x+a_0$, then
$$ \mbox{rev}(a)=\mbox{rev}_n(a)=a_0x^n+a_1x^{n-1}+\cdots+a_{n-1}x+a_0 $$
Equation (1) now reads
$$ \mbox{rev}_n(a)=\mbox{rev}_{n-m}(q)\cdot \mbox{rev}_m(b)+x^{n-m+1}\mbox{rev}_{m-1}(r),$$
and therefore,
$$\mbox{rev}_n(a)\equiv \mbox{rev}_{n-m}(q)\cdot \mbox{rev}_m(b)\, \mbox{mod}\, x^{n-m+1}.$$
Notice that $\mbox{rev}_m(b)$ has constant coefficient 1 and thus is invertible modulo $x^{n-m+1}$. Hence we find
$$ \mbox{rev}_{n-m}(q)\equiv \mbox{rev}_n(a)\cdot \mbox{rev}_m(b)^{-1} \, \mbox{mod}\, x^{n-m+1},$$
and obtain $q=\mbox{rev}_{n-m}(\mbox{rev}_{n-m}(q))$ and $r=a-q b$.

So now we have to solve the problem of finding, from a given $f\in D[x]$ and $l\in {N}$ with $f(0)=1$, a $g\in D[x]$ satisfying $fg\equiv 1\, \mbox{mod}\, x^l$.
If $l$ is a power of 2, then we can easily  obtain  the inversion by the following
iteration step
$$g_{i+1}=2g_i-fg_i^2  $$
In fact, if
$fg_i\equiv 1\, \mbox{mod}\, x^{2^i}$, then $x^{2^{i}}\, |\, 1-fg_i$, $x^{2^{i+1}}\, |\, (1-fg_i)^2$. Hence,
$x^{2^{i+1}}\, |\, 1-f(2g_i-fg_i^2)$.
Using the above iteration method, we have the following result:

\textbf{Theorem 1}.  \emph{Let $D$ be a ring (commutative, with 1), $f, g_0, g_1, \cdots, \in D[x]$, with $f(0)=1, g_0=1,$ and $g_{i+1}\equiv 2g_i-fg_i^2 \, \mbox{mod}\, x^{2^{i+1}}$, for all $i$. Then $fg_i\equiv 1\, \mbox{mod}\, x^{2^i}$ for all $i\geq 0$.}

By Theorem 1, we now obtain the following algorithm to compute the inverse of $f$\, mod\,$x^l$. We denote by log the binary logarithm.

\begin{tabular}{l}
\qquad   Algorithm 1: Inversion using Newton iteration \\ \hline
Input:¡¡\, $f\in D[x]$ with $f(0)=1$, and $l\in N$.\\
Output: \, $g\in D[x]$ satisfying $fg\equiv 1\, \mbox{mod}\, x^l$.\\
\qquad 1. $g_0\leftarrow 1, r\leftarrow \lceil \mbox{log}\, l\rceil$\\
\qquad 2. \textbf{for} $i=1, \cdots, r$ \textbf{do} $g_i\leftarrow (2g_{i-1}-fg_{i-1}^2) \, \mbox{rem}\, x^{2^i}$\\
\qquad 3. \textbf{Return} $g_r$  \\
  \hline
\end{tabular} \vspace*{5mm}

From the algorithm 1, one can easily obtain the following.\vspace*{5mm}

\begin{tabular}{l}
\qquad \qquad\qquad  Algorithm 2: Fast division with remainder \\ \hline
Input:\ ¡¡$a, b\in D[x]$, where $D$ is a ring (commutative, with 1) and $b\neq 0$ is monic.   \\
Output:\  $q, r\in D[x]$ such that $a=qb+r$ and deg\,$r<$ deg\,$b$.\\
\qquad 1. \textbf{if} deg\,$a<$\,deg\,$b$ \textbf{then return} $q=0$ and $r=a$ \\
\qquad 2. $m\leftarrow \mbox{deg}\,a-\mbox{deg}\,b$ \\
\qquad \ \textbf{call} Algorithm 1 to compute the inverse of $\mbox{rev}_{\mbox{deg}\,b}(b)\in D[x]$ modulo $x^{m+1}$\\
\qquad 3. $q^*\leftarrow \mbox{rev}_{\mbox{deg}\,a}(a)\cdot \mbox{rev}_{\mbox{deg}\,b}(b)^{-1}\, \mbox{rem}\, x^{m+1}$ \\
\qquad 4. \textbf{return} $q=\mbox{rev}_m(q^*)$ and $r=a-bq$\\
  \hline
\end{tabular}

\section{On the form of $l$ }

The authors \cite{GG03} stress that ``
if $l$ is not a power of 2, then the above algorithm computes too many coefficients of the inverse."
 They suggest  to compute the inverse modulo
  $  x^{\lceil l/2^r\rceil}, x^{\lceil l/2^{r-1}\rceil}, \cdots,$ $ x^{\lceil l/2\rceil}, x^ l$.
  For example,  suppose $l=11$, then
$x^{\lceil 11/2^4\rceil}=x$, $x^{\lceil 11/2^3\rceil}=x^2$, $x^{\lceil 11/2^2\rceil}=x^3$,  $x^{\lceil 11/2\rceil}=x^6$. In such case, one has to compute  $f^{-1}$ modulo
$  x, x^2, x^3, x^6, x^{11}$.
   It should be stressed that the authors  did not specify the iterative step.
   More serious, the sequence $1, 2, 3, 6, 11$ does not form an addition chain \cite{K97}.  Given a chain $\{a_i\}$ and $f$,  we can define the following
   iterative step
   $$ g_{a_k} \equiv g_{a_i}+g_{a_j} -fg_{a_i}g_{a_j}\, \mbox{mod}\, x^{a_k}, \ \mbox{if}\, a_k=a_i+a_j $$

  In fact, the suggestion is somewhat misleading. If $l$ is not a power of 2, the original algorithm 1 can be used to compute the inverse modulo
  $x^l$ without any additional cost. It suffices to observe the following fact.\\
   \centerline   {\textbf{Fact 1.} \emph{If $0<l\leq t$ and $x^t\,|\, 1-fg$, then $x^l\,|\, 1-fg$.}  }
The above fact is directly based on the divisibility characteristic. Based on the fact, we  obtain the following algorithm.

\begin{tabular}{l}
\quad   Algorithm 3: Inversion using   divisibility characteristic\\ \hline
Input:¡¡\, $f\in D[x]$ with $f(0)=1$, and $l\in N$.\\
Output: \, $g\in D[x]$ satisfying $fg\equiv 1\, \mbox{mod}\, x^l$.\\
\qquad 1. $g_0\leftarrow 1, r\leftarrow \lceil \mbox{log}\, l\rceil$\\
\qquad 2. \textbf{for} $i=1, \cdots, r-1$ \textbf{do} $g_i\leftarrow g_{i-1}\cdot(2-f\cdot g_{i-1}) \, \mbox{rem}\, x^{2^i}$\\
\qquad 3.  $g_r\leftarrow  g_{r-1}\cdot(2-f\cdot g_{r-1}) \, \mbox{rem}\, x^{l}$ \\
\qquad 4. \textbf{Return} $g_r$  \\
  \hline
\end{tabular} \vspace*{5mm}

\emph{Correctness.}  It suffices to observe that $l\leq 2^r$ where $r={\lceil \mbox{log}\, l\rceil}$.
 Hence $x^l\,|\, x^{2^{r}}$. Since $x^{2^r}\,|\, 1-f(2g_{r-1}-fg_{r-1}^2) $, we have $x^l\,|\, 1-f(2g_{r-1}-fg_{r-1}^2) $.
  That means $g_r$ is the inverse of $f$ modulo $x^l$, too.

\section{On the cost analysis }

To make a sound cost analysis, we need the following definition of multiplication time and its properties.

\textbf{Definition 1.} \emph{Let $R$ be a ring (commutative, with 1). We call a function $M: N_{>0} \rightarrow R_{>0}$ a multiplication time for $R[x]$ if polynomials in
$R[x]$ of degree less than $n$ can be multiplied using at most $M(n)$ operations in $R$. Similarly, a function $M$ as above is called a multiplication time for $Z$ if two
integers of length $n$ can be multiplied using at most $M(n)$ word operations. }

For convenience, we will assume that the multiplication time satisfies
$$M(n)/n \geq M(m)/m\ \mbox{if}\, n\geq m, \ \ M(mn)\leq m^2M(n),  $$
for all $n, m\in N_{>0}$. The first inequality yields the superlinearity properties
$$M(mn)\geq mM(n), \ M(m+n)\geq M(n)+M(m), \ \mbox{and}\, M(n)\geq n $$
for all $n, m\in N_{>0}$.

By the above  definition and properties, the authors obtained  the following result \cite{GG03}.

\textbf{Theorem 2.}
\emph{Algorithm 1 correctly computes the inverse of $f$ modulo $x^l$. If $l=2^r$ is a power of 2, then it uses at most $3M(l)+l\in O(M(l))$ arithmetic operations in $D$.}

\emph{Proof.}  In step 2, all powers of $x$ up to $2^i$ can be dropped, and since
$$g_i\equiv g_{i-1}(2-fg_i)\equiv g_{i-1} \, \mbox{mod}\, x^{2^{i-1}}, \eqno(2)$$ also the powers of $x$
less than $2^{i-1}$. The cost for one iteration of step 2 is $M(2^{i-1})$ for the computation of $g^{2}_{i-1}$, $M(2^i)$ for the product $fg_{i-1}^2\, \mbox{mod}\, x^{2^i}$,
 and then the negative of the upper half of $fg_{i-1}^2$ modulo $x^{2^i}$ is the upper half of $g_i$, taking $2^{i-1}$ operations. Thus we have
 $M(2^i)+M(2^{i-1})+2^{i-1}\leq \frac 3 2M(2^i)+2^{i-1}$ in step 2, and the total running time is
 $$ \sum_{1\leq i\leq r}\left(\frac 3 2M(2^i)+2^{i-1} \right)\leq \left(\frac 3 2 M(2^r)+2^{r-1}  \right)\sum_{1\leq i\leq r}2^{i-r}<3M(2^r)+2^r=3M(l)+l, \eqno(3)$$
 where we have used $2M(n)\leq M(2n)$ for all $n\in N$.  

There is a typo and an error in the above proof and theorem. \begin{itemize}

 \item{} In the above argument there is a typo (see Eq.(2)).

 \item{}The  cost for one iteration of step 2 is $M(2^{i})$ for the computation of $g^{2}_{i-1}$ instead of the original $M(2^{i-1})$, because it is computed under the module $x^{2^i}$, not $x^{2^{i-1}}$. Since the upper half of $f(g_{i-1}^2)$ modulo $x^{2^i}$ is the same as $g_i$ and the lower half of $g_i$ is the same as $g_{i-1}$, the cost for  the computation of $f(g_{i-1}^2)$ modulo $x^{2^i}$ only needs $M(2^{i-1})$.
      Therefore, according to the original argument  the bound should be
      $$ \sum_{1\leq i\leq r}\left( \frac 3 2 M(2^i)+2^{i-1} \right)\leq \left(\frac 3 2 M(2^r)+2^{r-1}  \right)\sum_{1\leq i\leq r}2^{i-r}<3M(2^r)+2^r\leq 12M(l)+2l, \eqno(4)$$
     The last estimation comes from  $ l\leq 2^r \leq 2l$.
 \end{itemize}

  Now, we make a formal cost analysis of algorithm 3.

  \textbf{Theorem 3.}
\emph{Algorithm 3 correctly computes the inverse of $f$ modulo $x^l$. It uses at most $5M(l)+l\in O(M(l))$ arithmetic operations in $D$.}

\emph{Proof.} The cost for step 2 is $3M(2^{r-1})+2^{r-1}$ (see the above cost analysis). The cost for step 3 is bounded by $2M(l)$. Since  $ 2^{r-1}\leq l\leq 2^r $,
the total cost is  $5M(l)+l$.

\section{Conclusion}

In this note, we revisit the fast division algorithm using Newton iteration. We show that the original Newton iterative step can be still used for any arbitrary exponent $l$ without the restriction that $l$ should be the power of 2. We also make a formal cost analysis of the method.
We think the new presentation is helpful  to grasp the  method entirely and deeply.

 \emph{Acknowledgements}
 We thank the National Natural Science
Foundation of China (Project 60873227), and the Key Disciplines of
 Shanghai Municipality (S30104).

\end{document}